**Understanding Ideality Factor in Wide-Band Gap Semiconductor Devices with Strong Recombination Planes**

Iris Celupica-Liu[1,a)], Tanay Tak[1], Yuh-Renn Wu[2], James S. Speck[1]

[1]Materials Department, University of California, Santa Barbara, CA 93106

[2]Graduate Institute of Photonics and Optoelectronics, Department of Electrical Engineering, National Taiwan University, Taipei 10617, Taiwan

[a)]Author to whom correspondence should be addressed: icliu@ucsb.edu

## Abstract

The ideality factor *n* has long been used to quantify how closely a diode's current-voltage characteristics follows the ideal drift diffusion model. In the classic diode equation, the ideality factor is a number than ranges between 1 and 2, with $n = 1$ indicating an ideal diode and $n = 2$ indicating that all current in the device is going towards an idealized defect recombination process in the depletion region. While this model gives a basic illustration of the effects of carrier recombination on diode current-voltage characteristics, it fails to explain experimentally observed ideality factors that are greater than 2. Using a *p-n* diode structure, modified to include a recombination plane in the depletion region, we devise a new model for calculating the recombination current. This new model indicates that depending on where the maximum recombination plane is located within the depletion region of the diode, its ideality factor only has a lower bound of 1 with no upper bound. Case studies of GaN diodes using this model yield results that are in better agreement with experimental measurements. and have the possibility of determining which carrier capture step is rate-limiting within a defect-assisted recombination cycle.

## Introduction

The ideality factor $n$ of a diode is a parameter used to gauge the degree of recombination occurring within the depletion region compared to the quasi-neutral region. It was defined to range between 1 and 2 by Sah, Noyce, and Shockley (SNS) in 1957 [1] through analysis of diode current-voltage characteristics; an ideality factor of $n = 1$ indicates a diode in which all recombination is happening in the quasi-neutral region (only drift-diffusion in the depletion region), while an ideality factor of $n = 2$ indicates that all current is participating in recombination-generation within the depletion region. The ideality factor, $n$, is determined by fitting the current density-voltage ($J - V$) characteristics of diodes to $J = J_0 \exp\left(\left(\frac{qV}{nkT}\right) - 1\right)$, where $J_0$ is the reverse bias saturation current, $q$ is the fundamental charge, $k$ is the Boltzmann constant, and $T$ is the junction temperature.

It is important to note that in the original SNS analysis, the only recombination mechanism that was considered was Shockley-Read-Hall (SRH) recombination, taking place at traps located within the depletion region of the device and maximized where $n = p$ [2], [3]. While the SNS model has adequately described nonradiative recombination for Si- and Ge-based diodes, it must be revisited for semiconductors in which there is fast radiative recombination or localized regions of recombination (e.g. QWs, growth interrupts, regrowth interfaces) – specifically gallium nitride (GaN)-based p-(i-)n diodes [4], [5].

Ideality factors of up to 9 have been measured for GaN-based diodes[4], [5], [6], [7]. Analyzing AlGaN/GaN p-n junctions, Shah et al. considered the multiple junctions within the diode as the source of ideality factors exceeding a value of 2, with some GaN p-n junctions attaining values as high as 6.9. Yet, the authors do not consider the physical significance of the SNS ideality factor and the role recombination plays in determining ideality factor [5]. Additionally, Fedison *et al.* [4] observe ideality factors between 7 to 9 in GaN p-n diodes, positing that such high ideality factors may arise from the presence of deep levels. The authors, however, do not further explain the mechanism by which these high ideality factors may arise. Thus, while these studies have formulated models or postulates aimed at explaining anomalously high ideality factors in GaN-based diodes, these explanations remain incomplete by not quantitatively accounting for the effect of carrier recombination on diode forward characteristics.

Beyond power electronics, GaN has been indispensable for the development of efficient visible-wavelength light-emitting diodes (LEDs). Like p-n junctions, LEDs have also been observed to have ideality factors exceeding 2 [6], [7]. Many studies have attributed these higher ideality factors to be related to trap-assisted tunneling (TAT) which leads to a tunneling current in the device [4], [6]. Auf der Maur et al. (2014) performed a comparison between TAT simulation results and measurements done on InGaN/GaN LEDs with ideality factors ranging from $n = 4$ to $n = 14$ [6]. The authors concluded that the TAT mechanism may provide plausible explanation of ideality factors >2 for biases below ~2.5 V, only if Mg used in the electron blocking layer (EBL) is treated as a deep level. On the contrary, Masui [7] posits that high ideality factors in LEDs arise from effects relating to minority carrier partitioning relative to the placement of the InGaN quantum well (QW). However, the author dismisses the existence of impurities and the role of defect-assisted nonradiative recombination in the active region, which is not in line with what has been experimentally [8] and computationally [9] observed. Thus, a more complete model for explaining high observed ideality factors in LEDs is still needed, though it is beyond the scope of the present work, which only focuses on understanding the relationship between ideality factor and defect-assisted recombination in GaN homostructure p-i-n diodes. Use of homostructure devices eliminates complications relating to the inclusion of InGaN QWs (such as polarization barriers, localized carrier accumulation and depletion, alloy disorder, etc.) that would need to be accounted for when constructing a complete model for LEDs.

This work focuses on constructing a new model of diode ideality factor by analyzing a GaN-based p-i-n homojunction containing a thin layer of impurity-doped material known to create deep levels within the band gap, located in the depletion region. While this analysis is based on the original SNS and SRH analyses of diode J-V characteristics and nonradiative recombination, parameters have been adjusted to account for GaN's larger band gap and the incomplete ionization of Mg in p-type GaN material. Additionally, it will be shown that this model can be used to study mechanisms of recombination in a way that is not previously possible by probing the rate-limiting carrier capture step within a defect-assisted recombination cycle (DARC) [10]. This model may be modified to include the effects of trap-assisted Auger-Meitner recombination (TAAR), opening the possibility to study a different but important source of nonradiative recombination.

**Methods**

Rather than focusing on an idealized p-n junction, this model is general enough to account for the inherent asymmetry of the depletion region in a GaN-based p-i-n diode, with a thin plane of nonradiative centers incorporated into the depletion region at some arbitrary position. This plane dominates the recombination in the diode for applied bias well below the built-in voltage and behaves similarly to how a quantum well functions in terms of recombination within an LED [7] or impurity planes introduced into a diode structure during growth due to growth interrupts or at regrowth interfaces.

To further simplify the following analysis, several reasonable assumptions are made. First, it is assumed that the thickness of the impurity plane is negligible compared to the total length of the depletion region. This permits the assertion that parameters relevant to recombination (carrier concentrations) are constant within the recombination plane, which simplifies mathematical analysis. It is easy to fulfill this assumption, as it just requires sufficient doping of the n- and p-regions of the device, such that an "i"-region of sufficient length can be fully depleted. Dimensions of an example device considered in this model are given in **Figure 1**. Additionally, to derive expressions that no longer depend on carrier quasi-Fermi levels, which are difficult to compute analytically, we observe that the electron and hole quasi-Fermi levels nearly coincide with the conduction band edge (CBE) and valence band edge (VBE), in their respective quasi-neutral regions, in GaN. This is illustrated using 1D Schrodinger-Poisson drift-diffusion, using simulation software from [11] to simulate an identical diode, except without the impurity plane (**Figure 2**). Parameters used in the simulation are given in **Table I**. Finally, we assume that the nonradiative centers being created within the junction are deep ($> 3kT$ from the band edges), as SRH theory predicts that deeper trap levels act as more efficient nonradiative recombination centers [2].

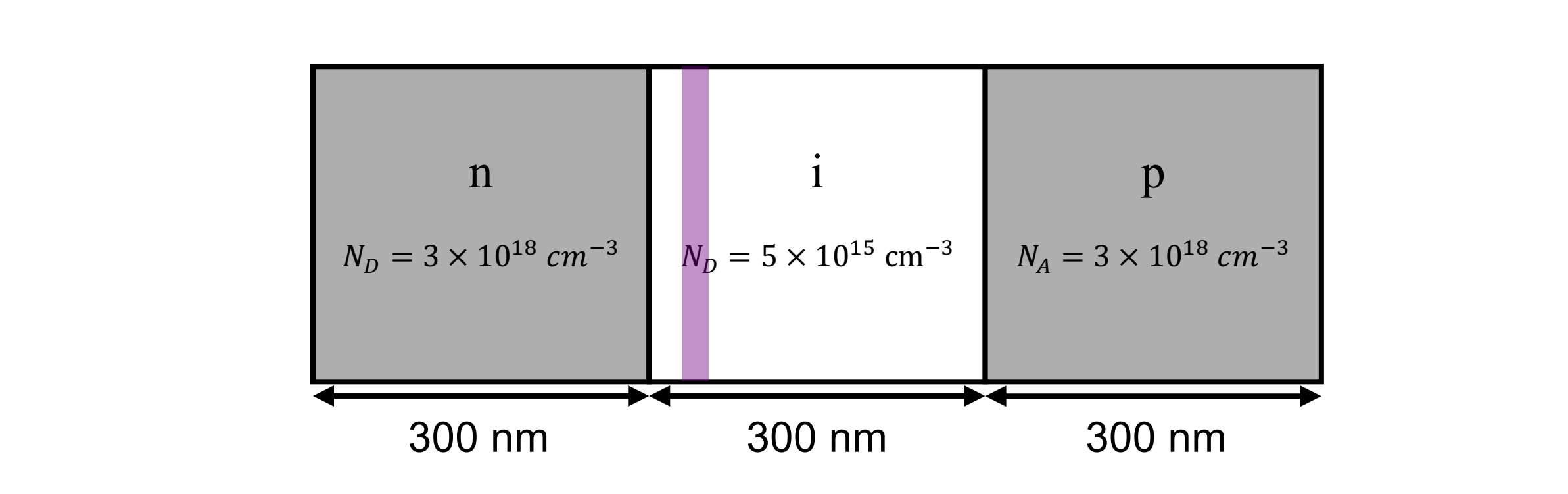

**Figure 1**: Schematic diagram of the device structure considered and simulated. Dopant concentrations are given for the n- and p-regions, while approximate background carrier concentration is given for the unintentionally doped (UID) "i"- region. The impurity plane is illustrated by the purple shaded region.

The net recombination rate $R$ from defect-mediated recombination was derived by Shockley, Read, and Hall in their 1952 publications [2], [3] and is given by

$$R = \frac{np - n_i^2}{\tau_n(p + p_1) + \tau_p(n + n_1)}, \tag{1}$$

with $n_1 = n_i \exp\left(\frac{E_t' - E_i}{kT}\right)$, $p_1 = n_i \exp\left(\frac{E_i - E_t'}{kT}\right)$, $n = n_i \exp\left(\frac{E_{Fn} - E_i}{kT}\right)$, and $p = n_i \exp\left(\frac{E_i - E_{Fp}}{kT}\right)$. $n$ and $p$ are the free electron and hole concentrations, respectively; $n_i$ is the intrinsic carrier concentration; $\tau_n$ and $\tau_p$ are the minority electron and hole lifetimes, respectively; $E_{Fn}$ and $E_{Fp}$ are the electron and hole quasi-Fermi levels, respectively; and $E_i$ is the intrinsic energy level.

| Material | Doping [$cm^{-3}$] | $E_a$ [meV] | $e^-/h^+$ mobility [$cm^2$/Vs] | $\tau_{n,p}$ [ns] |
|---|---|---|---|---|
| p-GaN | $3 \times 10^{20}$ | 200 | 300/10 | 200 |
| UID | $5 \times 10^{15}$ | 20 | 300/10 | 200 |
| n-GaN | $3 \times 10^{18}$ | 20 | 300/10 | 200 |

**Table I**: Parameter settings for the simulated GaN p-i-n diode. p-GaN corresponds to epi surface.

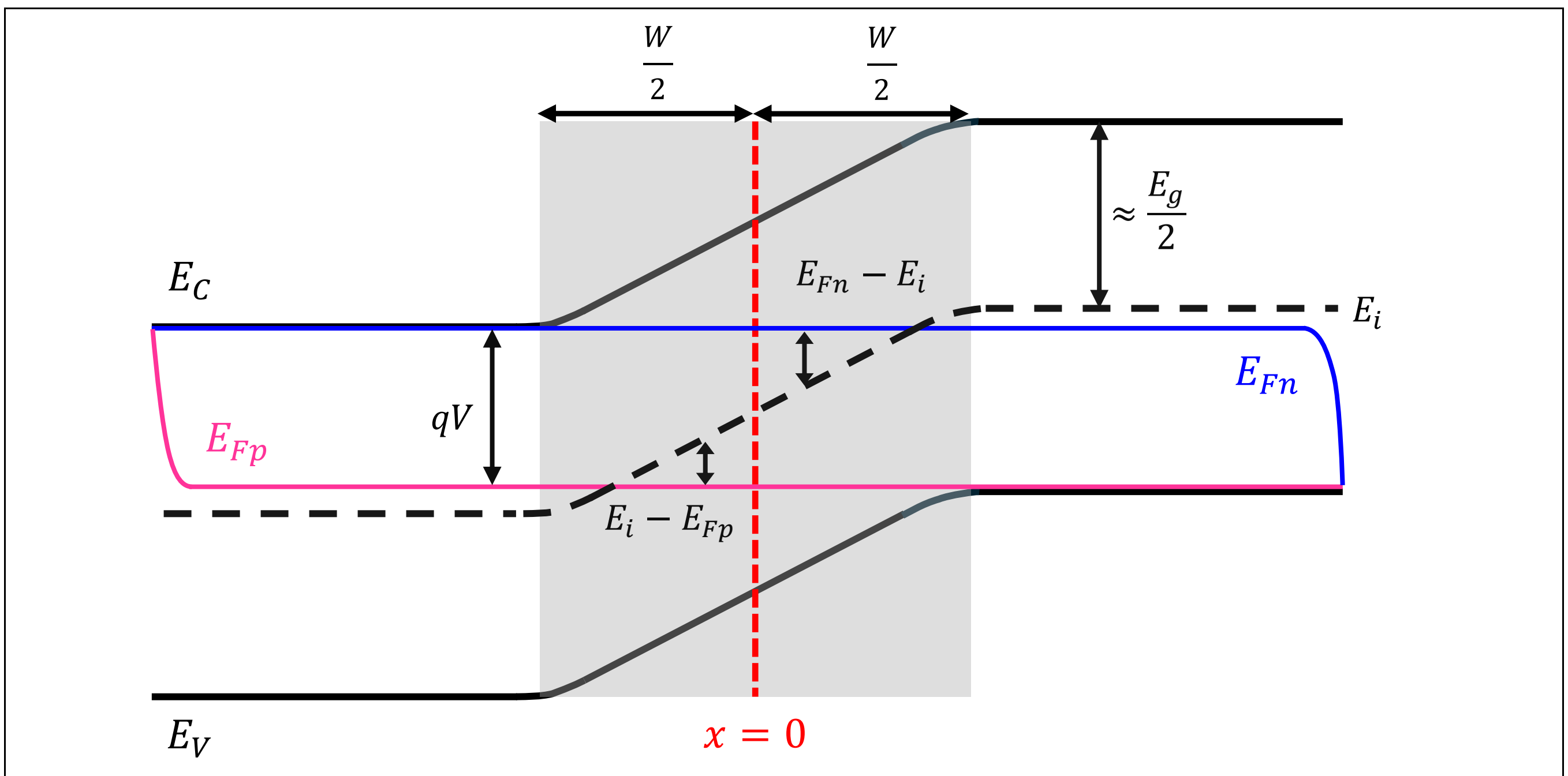

**Figure 2**: Schematic band diagram of a GaN p-i-n junction at forward bias. The carrier quasi-Fermi levels are approximately coincident with the band edges in the quasi-neutral regions. Origin of the coordinate system for the following model developed was set to be at the midpoint of the depletion region for all bias values. Doping values for the structure modeled in this schematic are given in **Table I**.

Due to the dependence of $n_1$ and $p_1$ on the intrinsic carrier concentration $n_i$, they are negligible because $n_i$ is vanishingly small in wide band gap semiconductors, such as GaN that we consider ($\sim 10^{-10}$ cm$^{-3}$). It is worth noting that while this assumption breaks down for defect levels within 300 meV of the band edges ($n_1 \approx 2.5 \times 10^{15}$cm$^{-3}$ for a donor ~25 meV from the CB, $p_1 \approx 8 \times 10^{15}$cm$^{-3}$ for an acceptor 200 meV from the VB) due to the large value of the exponential term, most deep levels of interest are sufficiently far from the band edges such that $n_1$ and $p_1$ are dominated by $n_i$. Under forward bias, the product *np* is constant across the depletion region and $n_i^2$ vanishes. However, $n_i^2$ must be kept in the analysis so that there is continuity with zero-bias/equilibrium state. We begin with the most general case, which would be that for an impurity plane located at the center of a symmetrical p-i-n junction. For GaN, which usually has asymmetrical junctions due to the deep nature (~180 meV) of the common Mg and shallow nature of the common donor Si (~25 meV), and thus commonly $N_A \gg N_D$, this would be equivalent to placing the plane at the point where $n = p$. Under these assumptions, the net recombination rate becomes

$$R \approx \frac{np - n_i^2}{\tau_n p + \tau_p n}. \tag{2}$$

To derive the recombination current $J_R$, one must integrate the net recombination rate over the length in which recombination is occurring, before multiplying it by the fundamental unit of charge *q*. Since the impurity plane is meant to dominate recombination in the device, it can be assumed that recombination happening elsewhere in the device is negligible. As such, one only needs to integrate of the length of the plane, which is assumed to be sufficiently thin such that carrier concentrations remain approximately constant across it. The calculation proceeds as follows:

$$J_{R,center} = q \int_{\frac{W}{2}+x-\frac{t}{2}}^{\frac{W}{2}+x+\frac{t}{2}} \frac{np - n_i^2}{\tau_n p + \tau_p n} dx = \frac{qnpt - qn_i^2 t}{\tau_n p + \tau_p n}, \tag{3}$$

with *x* representing the location of the center of the impurity plane relative to the center of the depletion region $\frac{W}{2}$, *t* is the thickness of the impurity plane, and *W* is the total length of the depletion region. Here, the origin of the coordinate system being used is taken as the center of the total depletion width. Note that for an asymmetric junction, this will not correspond to half the total length of the diode and will move as applied bias is varied. In the case of a GaN-based diode, in which the p-type region tends to be more heavily doped than the n-type region, the origin will start slightly towards the p-side of the diode, as depletion into the n-type region will be longer than that into the p-type region. As applied bias is increased and depletion into the intentionally doped regions is decreased, the origin will start to move closer to the exact center of the diode. This is shown in **Figure 3**, exaggerated to better illustrate how the origin changes with applied bias.

From basic device physics, the total drift-diffusion current $J_{ideal}$ in a p-i-n junction is

$$J_{ideal} = qn_i^2\left[\frac{D_N}{L_N N_A} + \frac{D_P}{L_P N_D}\right]\left[e^{\frac{qV}{kT}} - 1\right], \tag{4a}$$

in which $D_N$ and $D_P$ are the diffusion coefficients of electrons and holes, and $L_N$ and $L_P$ are the diffusion lengths for electrons and holes, respectively. $N_A$ and $N_D$ are the acceptor and donor densities, respectively, and $V$ is the applied bias. This expression, on its own, represents the ideal diode equation, which states that an ideal diode only has current due to drift and diffusion processes. To realize the total diode current, one simply needs to take the sum of the drift-diffusion and recombination currents:

$$J_{tot} = J_{ideal} + J_{R,center} = qn_i^2\left[\frac{D_N}{L_N N_A} + \frac{D_P}{L_P N_D}\right]\left[e^{\frac{qV}{kT}} - 1\right] + \frac{qnpt - qn_i^2 t}{\tau_n p + \tau_p n} \tag{4b}$$

or

$$J_{tot} = qn_i^2\left[\frac{D_N}{L_N N_A} + \frac{D_P}{L_P N_D}\right]\left[e^{\frac{qV}{kT}} - 1\right] + \frac{tqn_i e^{\frac{eV}{kT}} - qn_i t}{\tau_n e^{\frac{E_i - E_{Fp}}{kT}} + \tau_p e^{\frac{E_{Fn} - E_i}{kT}}}, \tag{4c}$$

since $np = n_i^2 e^{\frac{qV}{kT}}$. However, before diode turn-on, the total current is dominated by recombination such that

$$J_{tot} \approx J_{R,center} = \frac{tqn_i e^{\frac{qV}{kT}} - qn_i t}{\tau_n e^{\frac{E_i - E_{Fp}}{kT}} + \tau_p e^{\frac{E_{Fn} - E_i}{kT}}}.$$

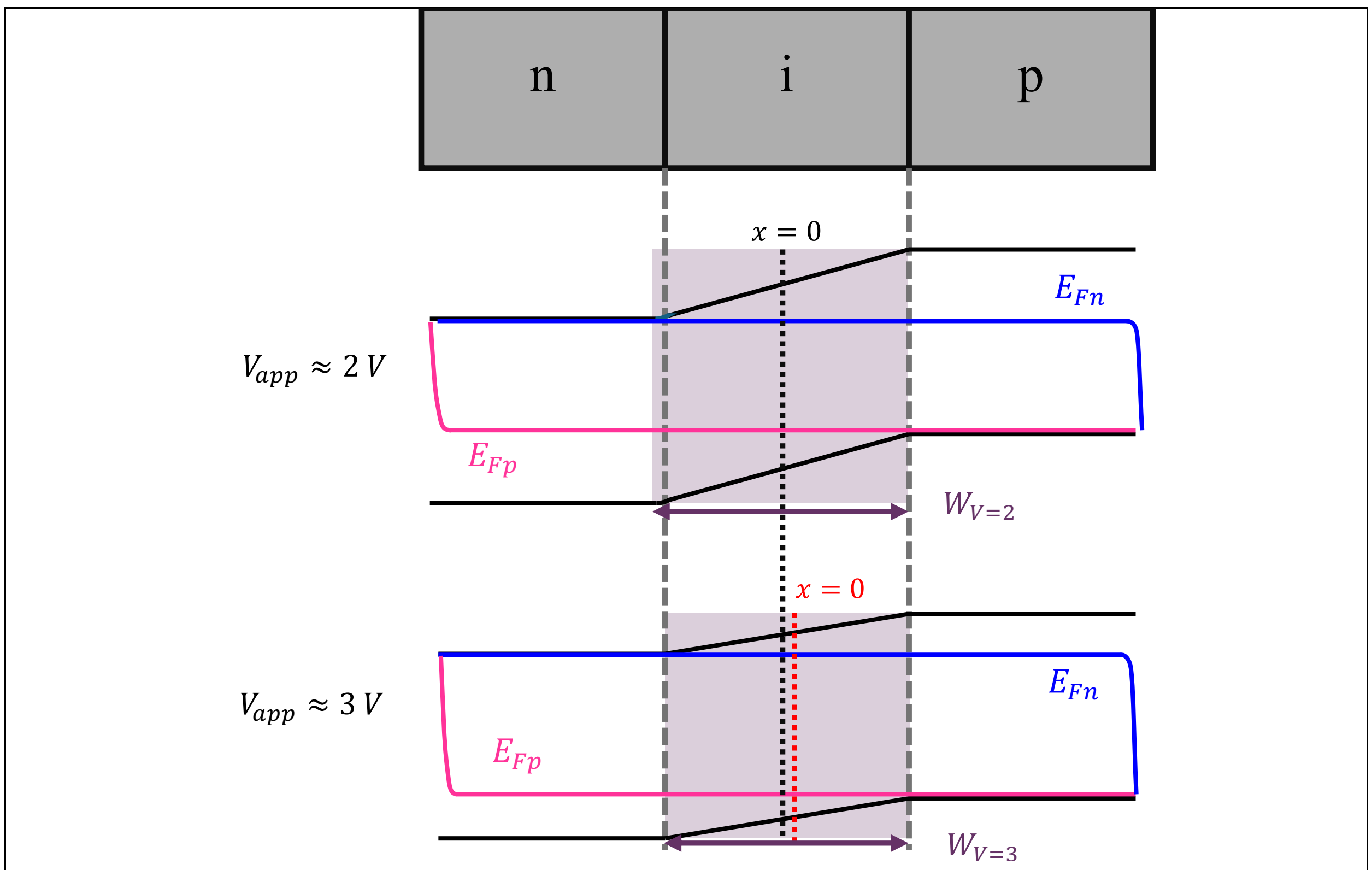

**Figure 3**: Schematic band diagram of the p-i-n structure for two values of applied bias, illustrating the slight rightward shift of the origin, which we define as the location $n = p$, with increasing bias.

Unlike previous works that explore ideality factor – which focus only on explaining experimental results using revised models – the structure proposed here can be used to enhance the understanding of defect-assisted recombination cycles (DARCs). DARCs refer to the various mechanisms of defect-assisted nonradiative recombination, such as Shockley-Read-Hall (SRH) recombination and trap-assisted Auger-Meitner recombination (TAAR). Though SRH and TAAR involve a different number of carriers and dominate at different current injection regimes, they nonetheless both focus on the recombination of an electron and a hole that is mediated by a defect level within the band gap. This work will focus on understanding SRH processes at low applied biases, though the model's application to TAAR processes will be explored in later experimental and theoretical works.

To reiterate, a DARC is a model used to conceptualize how recombination occurs at a given defect by accounting for all known possible carrier capture mechanisms. Shockley and Read were the first to postulate that recombination only takes place if both a hole and an electron are captured at the defect [2]. Theoretically, a given defect can capture and emit carriers in an indefinite number of cycles, but only the capture of both a hole and an electron will lead to recombination. This is most likely when the electron and hole capture rates are equal, which is not always the case in real defect-semiconductor system. Though fundamental, Shockley and Read's model is limited in scope, as they only consider SRH recombination, which has been hypothesized to occur via the multi-phonon emission (MPE) mechanism. Presently, however, we now understand that recombination can occur by many carrier capture mechanisms, including but not limited to: band-to-band recombination; multi-phonon emission; trap-assisted Auger-Meitner recombination (TAAR); and radiative recombination. Carriers can be captured into the defect level via any of these mechanisms, indicating that a full recombination cycle can result from numerous combinations of carrier capture mechanisms. Recent publications by Tak *et al.* [10] and Zhao *et al.* [12] more aptly capture the recombination physics by accounting for band-to-band, radiative, MPE-assisted, and trap-assisted Auger-Meitner recombination. However, neither propose experimental avenues by which to test the models.

To study the specific carrier capture steps involved in a DARC, the impurity plane can simply be placed at different locations within the diode's depletion region. If the plane is placed closer to the n-type side of the diode, there will be an extreme deficit in holes that are available for recombination (**Figure 4**). With electrons greatly outnumbering holes, it will be more difficult to complete a recombination cycle unless hole capture is sufficiently fast. Similarly, if the impurity plane is placed closer to the p-type side of the diode, the recombination cycle will be difficult to complete unless electron capture is sufficiently fast. Of course, there is additional consideration for the difference between $\tau_n$ and $\tau_p$ of the defect that also needs to be considered. As will be shown below, the ideality factor will change depending on which carrier capture step is rate-limiting, illustrating how this model can be applied to better understand defect-assisted recombination.

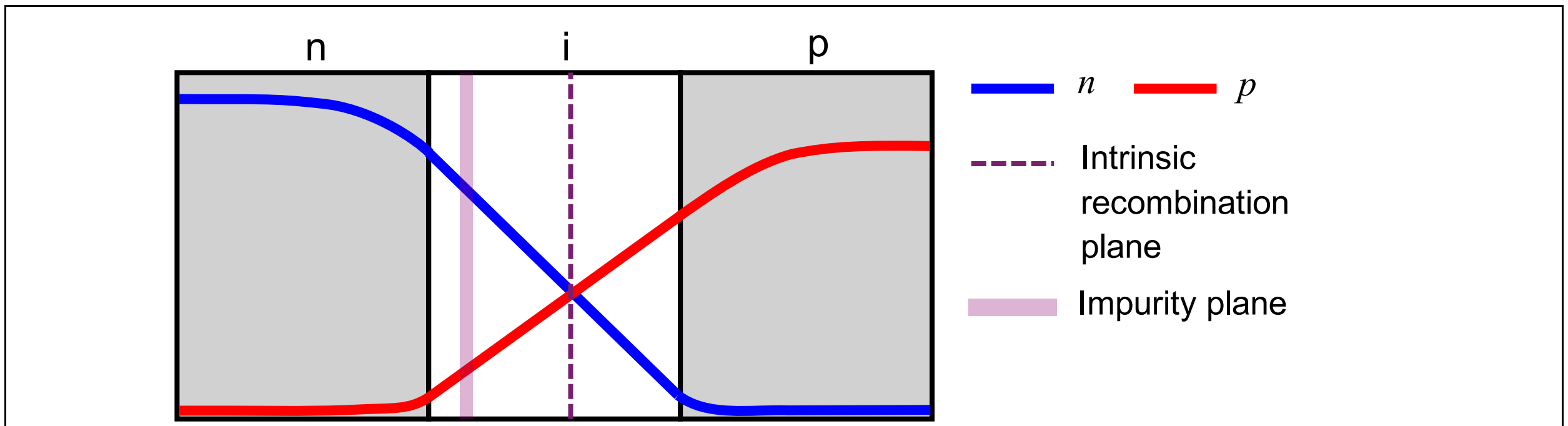

**Figure 4**: Schematic diagram of device structure with hole (red) and electron (blue) concentrations overlaid. Placing the impurity plane (pink shaded) closer to the n-side limits the concentration of holes available for recombination.

Recall, that the net recombination rate $R$ was said to be equal to $\frac{np-n_i^2}{\tau_n p+\tau_p n}$ for the conditions of interest in this work. In scenarios when $\tau_n p \gg \tau_p n$, the expression can be simplified further:

$$R = \frac{np - n_i^2}{\tau_n p + \tau_p n} \approx \frac{np - n_i^2}{\tau_n p}.$$

Physically, this can occur when the plane of maximum recombination is closer to the p-side of the junction where $p \gg n$ when $\tau_n \approx \tau_p$. Otherwise, it may also occur when the plane of maximum recombination is closer to the n-side of the junction if it is composed of defects whose $\tau_n \gg \tau_p$. Then, the total diode current density for low applied biases becomes:

$$J_{tot,p-side} = \frac{tqn_i^2 e^{\frac{qV}{kT}} - qn_i^2 t}{\tau_n n_i e^{\frac{E_i - E_{Fp}}{kT}}} = \frac{qtn_i}{\tau_n}(e^{\frac{qV}{kT}} - 1)e^{\frac{E_{Fp}-E_i}{kT}} \approx \frac{qtn_i}{\tau_n} e^{\frac{qV}{kT}} e^{\frac{E_{Fp}-E_i}{kT}} \tag{5}$$

From the band diagrams of a GaN p-i-n junction, we can rewrite $\frac{E_{Fp}-E_i}{kT}$ in a more tractable form by assuming that the electron and hole quasi-Fermi levels coincide with the conduction band and valence band edges, respectively (**Figure 2**). This assumption results in $E_i - E_{Fp} \approx \frac{E_g}{2}$ in the p-type quasi-neutral region, which differs from the real solution by < 0.1 eV since $E_i$ in GaN is roughly 1.75 eV above $E_V$ and $E_{Fp}$ is at most ~0.15 eV above the $E_V$ depending on doping. Note that the origin of the coordinate system used to arrive at these expressions will shift closer to the p-type region as applied bias increases, corresponding to the decreasing depletion depth into the n-type region; this is shown schematically in **Figure 3**. Using these assumptions, the difference $E_i - E_{Fp}$ increases linearly as one moves toward the p-type region. This is represented by

$$E_i - E_{Fp} = \Delta E(x, V) = mx + b, \text{with } m = q\left(\frac{V_{bi} - V}{W}\right)$$

As a result,

$$\Delta E(x, V) = q\left(\frac{V_{bi}-V}{W}\right)x + \gamma \tag{6}$$

The boundary condition at $x = \frac{W}{2}$ can be used to determine $\gamma$ for an arbitrary applied bias $V$. For $x = \frac{W}{2}$, $\Delta E$ becomes

$$\Delta E\left(x = \frac{W}{2}, V\right) = q\left(\frac{V_{bi} - V}{W}\right)\frac{W}{2} + \gamma = \frac{E_g}{2}$$

and rearranging yields that

$$\gamma = \frac{E_g}{2} - q\left(\frac{V_{bi}-V}{W}\right)\frac{W}{2} = \frac{E_g}{2} - \frac{q}{2}(V_{bi} - V).$$

Substituting $\gamma$ back into (6), we find that

$$\Delta E(x, V) = q\left(\frac{V_{bi} - V}{W}\right)x + \frac{E_g}{2} - \frac{q}{2}(V_{bi} - V) = q(V_{bi} - V)\left(\frac{x}{W} - \frac{1}{2}\right) + \frac{E_g}{2}$$

Finally, to get the quantity of interest, $E_{Fp} - E_i$, the above expression simply needs to be multiplied by -1 to get

$$E_{Fp} - E_i = -\Delta E = -q(V_{bi} - V)\left(\frac{x}{W} - \frac{1}{2}\right) - \frac{E_g}{2} = -\frac{qV_{bi}x}{W} + \frac{qV_{bi}}{2} + \frac{qVx}{W} - \frac{qV}{2} - \frac{E_g}{2}$$
$$= \frac{-2qV_{bi}x + WqV_{bi} + 2qVx - WqV - WE_g}{2W}$$

Since $W$ will also depend on applied bias $V$ (as the amount of depletion into the n- and p-regions depends on applied bias), this will complicate the final computation and prevent the ideality factor from being written in a closed-form expression. However, due to the high concentration of dopants in the n- and p-regions relative to the i-region, depletion into the bulk regions will be negligible. Since the i-region is fully depleted at the low biases of interest for the structure considered here, then the size of the depletion width is approximately the width of the i-region itself. Thus, $W = W_i$ can be substituted into the above expression:

$$E_{Fp} - E_i = -\Delta E = \frac{-2qV_{bi}x + W_iqV_{bi} + 2qVx - W_iqV - W_iE_g}{2W_i}. \quad (7)$$

Substituting (6) back into (5), $J_{tot,p-side}$ becomes

$$J_{tot,p-side} = \frac{qtn_i}{\tau_n} e^{\frac{qV_{bi}(W_i-2x)-W_iE_g}{2W_ikT}} e^{\frac{qV(2x+W_i)}{2W_ikT}}. \quad (8a)$$

This expression can be normalized to $W_i$ by defining $\tilde{x} = \frac{x}{W_i}$ where $\tilde{x}$ is bounded by $\left[\frac{-1}{2}, \frac{1}{2}\right]$, i.e. positive values indicate the recombination plane is closer to the p-side of the junction and negative values indicate that the recombination plane is closer to the n-side of the junction. This allows (8a) to be rewritten as

$$J_{tot,p-side} \approx \frac{qtn_i}{\tau_n} (e^{\frac{qV_{bi}(W_i-2x)-W_iE_g}{2W_ikT}}) e^{\frac{qV(2\tilde{x}+1)}{2kT}}. \quad (8b)$$

From equation (8b), we find that the ideality factor $n$ depends only on $\tilde{x}$ and is defined as

$$n = \frac{2}{2\tilde{x} + 1}. \quad (9)$$

Thus, for the case of the impurity slab/recombination plane where $\tau_n p \gg \tau_p n$, the ideality factor $n$ will be bounded in the range $n \in [1, \infty)$. More specifically, if the recombination plane is at the center of the depletion region, the ideality factor will be 2, in agreement with the conventional SNS analysis. As the recombination plane approaches the i-p interface ($\tilde{x} \to \frac{1}{2}$), the ideality factor becomes 1, as shown in **Figure 5**. On the other hand, as the recombination plane approaches the n-i interface ($\tilde{x} \to \frac{-1}{2}$), the ideality factor diverges to infinity.

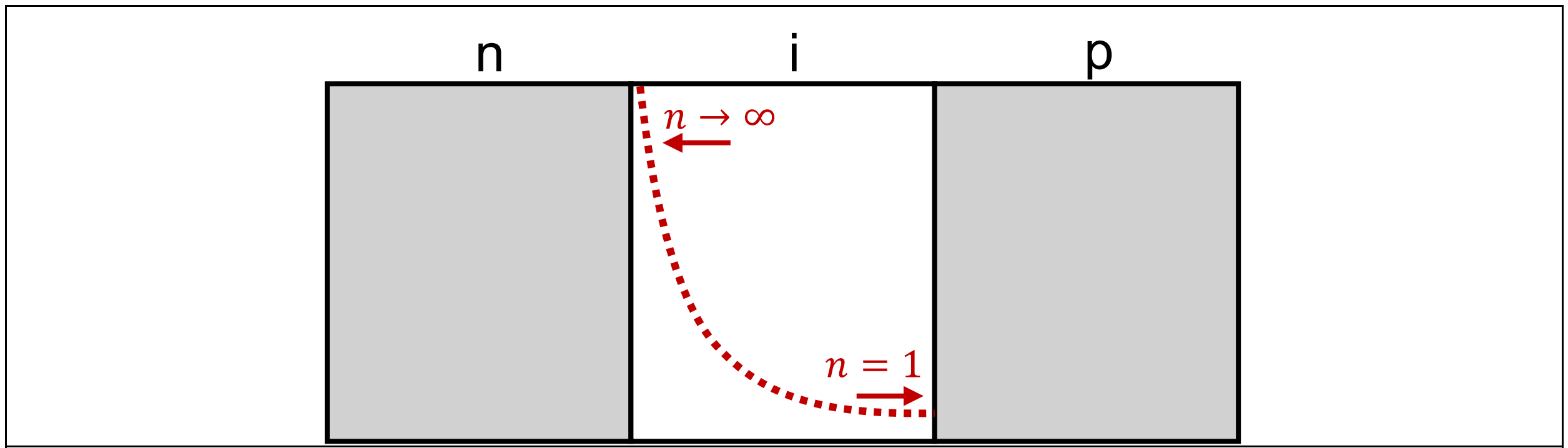

**Figure 5**: Schematic of how ideality factor $n$ (dashed red) changes as a function of position within the diode depletion region. $n$ approaches 1 at the i-p interface and diverges as the n-i interface is approached. This corresponds to electron-limited nonradiative recombination. Recall that $\tau_n p \gg \tau_p n$ in the recombination plane for this to hold true.

Before moving on, let us consider the physical picture of the results so far. Recall that this simplification, in which the $\tau_p n$ term in denominator is neglected, relies on an assumption that $\tau_n p \gg \tau_p n$, resulting in $R \approx \frac{n}{\tau_n}$. If the recombination plane is at $\tilde{x} = 0$, where $n = p$, then $\tau_n \gg \tau_p$ and the defect-assisted recombination occurs, however is limited by the slower electron capture step. The fact that recombination occurs here results in an ideality factor of 2, in agreement with the conventional SNS model. As the recombination plane approaches the i-p interface, $\tilde{x} \to \frac{1}{2}$ and $n \to 1$, the lack of electron density limits the recombination from occurring, regardless of whether $\tau_n \gg \tau_p$ or $\tau_n \approx \tau_p$. This reduces the recombination rate significantly compared to the $\tilde{x} = 0$ case, effectively meaning that as the recombination rate decreases as $\tilde{x} \to \frac{1}{2}$, where at $\tilde{x} = \frac{1}{2}$ there is no recombination and thus it is an ideal diode with an ideality factor of 1. On the other hand, as $\tilde{x} \to \frac{-1}{2}$ and $p \to 0$, the only way to maintain $\tau_n p \gg \tau_p n$ is by considering a defect with $\tau_n \gg \tau_p \frac{n}{p}$. For such a defect with many orders of magnitude slower electron capture rate to exist, it must be an acceptor in a negative charge state that repels electrons away from it, effectively behaving as a trap instead of a recombination center. However, recombination can still occur in this case, as $\tau_p$ being substantially smaller than $\tau_n$ implies that the hole capture step is incredibly fast, such that completion of the recombination cycle is completed by the slower electron capture step. Electron capture becomes possible because of the extreme excess of electrons in this part of the junction increases the capture rate.

An identical analysis can be done for the case where $\tau_p n \gg \tau_n p$ in the impurity slab. Then the total diode current density becomes

$$J_{tot,n-side} \approx \frac{qtn_i^2 e^{\frac{qV}{kT}} - qtn_i^2}{\tau_p n_i e^{\frac{E_{Fn}-E_i}{kT}}} = \frac{qtn_i}{\tau_p}(e^{\frac{qV}{kT}} - 1)e^{\frac{E_i-E_{Fn}}{kT}} \approx \frac{qtn_i}{\tau_p} e^{\frac{qV}{kT}} e^{\frac{E_i-E_{Fn}}{kT}}. \qquad (10)$$

As done for the above case, the difference $E_i - E_{Fn}$ can be written in a more tractable form using the band diagram and assumptions appropriate to GaN:

$$E_i - E_{Fn} = \Delta E'(x, V) = mx + b, \text{with } m = q\left(\frac{V_{bi} - V}{W}\right) \qquad (11)$$

In this case, we take $E_{Fn}$ to be the "zero"/reference energy level in this coordinate system. Applying the appropriate boundary conditions for the case of $x = \frac{W}{2}$, (10) becomes

$$\Delta E'(x, V) = E_i - E_{Fn} = q\left(\frac{V_{bi} - V}{W}\right)x + \frac{1}{2}\left[E_g - q(V + V_{bi})\right]. \qquad (12)$$

As mentioned above, it is reasonable assume that $W = W_i$, such that $J_{tot,n-side}$ becomes:

$$J_{tot,n-side} \approx \frac{qtn_i}{\tau_p}\left(e^{\frac{qV_{bi}(2x-W_i+W_iE_g)}{2W_ikT}}\right)e^{\frac{qV(W_i-2x)}{2W_ikT}}. \qquad (13a)$$

To normalize the expression, we again invoke $\tilde{x} \equiv \frac{x}{W_i}$ where $\tilde{x}$ is bounded by $\left[\frac{-1}{2}, \frac{1}{2}\right]$, i.e. positive values indicate the recombination plane is closer to the p-side of the junction and negative values indicate that the recombination plane is closer to the n-side of the junction. This allows that (12) can be rewritten as

$$J_{tot,n-side} \approx \frac{qtn_i}{\tau_p}\left(e^{\frac{qV_{bi}}{2kT}(2\tilde{x}-1+E_g)}\right)e^{\frac{qV(1-2\tilde{x})}{2kT}}. \qquad (13b)$$

Using the form of equation (12b), the ideality factor *n* for the case of the impurity plane with $\tau_p n \gg \tau_n p$ can be written in a closed-form expression as

$$n = \frac{2}{1 - 2\tilde{x}}. \qquad (14)$$

Like the earlier case, the ideality factor *n* will be bounded in the range $n \in [1, \infty)$. Ideality factor will be equal to 1 at the n-i interface and will begin to diverge to infinity as *x* approaches the i-p interface, as shown in **Figure 6**. Unlike the prior case, for the ideality factor to approach infinity, $\tau_p \gg \tau_n \frac{p}{n}$ towards the i-p interface, implying that the defect in this impurity plane must behave as a donor trap instead of an acceptor trap.

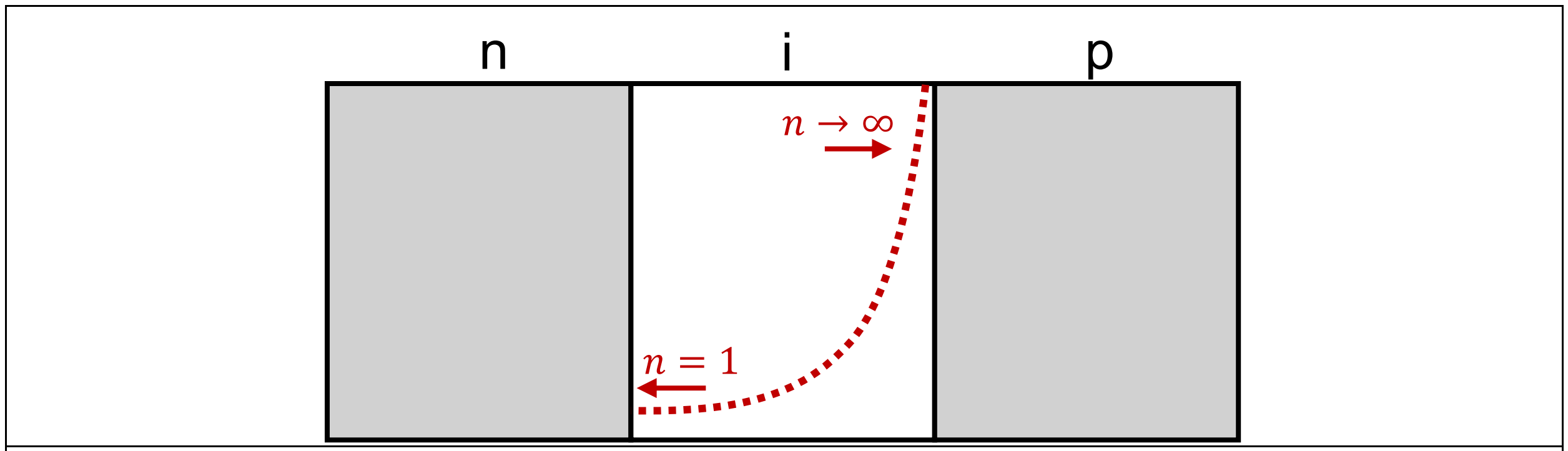

**Figure 6**: Schematic of how ideality factor n (dashed red) changes as a function of position within the diode depletion region. $n$ approaches 1 at the n-i interface and diverges as the i-p interface is approached. This corresponds to hole-limited nonradiative recombination. Recall that $\tau_p n \gg \tau_n p$ in the recombination plane for this to hold true.

**Case Study: Predicted J-V curves of GaN p-i-n with C-doped plane**

Carbon is a common atmospheric contaminant in both metal organic chemical vapor deposition (MOCVD) and molecular beam epitaxy (MBE) reactors, often observed at regrowth interfaces and growth interruptions in higher concentrations ($\sim 5 \times 10^{20}$ cm$^{-3}$) and as a background impurity in lower concentrations ($\sim 1 \times 10^{16}$ cm$^{-3}$), though exact concentrations are reactor dependent. While there is evidence that C possesses amphoteric behavior depending on whether it substitutes onto the Ga site or N site, only the $C_N$ defect will be considered here, as N appears to be the dominant substitutional site in MOCVD- [13] and MBE-grown [14] material. The $C_N$ defect with (-1/0) charge states, with an energy level of $E_V + 1.06$ eV in the band gap, is believed to contribute to nonradiative recombination in GaN-based devices [12], [13]. Multiple studies [13], [14], [15] have observed an increase in compensation of free electrons as carbon concentration is increased. This suggests the tendency of carbon to act as an acceptor trap, despite its closer proximity to the valence band. This electron trapping behavior is also somewhat unexpected, given an electron capture coefficient of $C_n = 3 \times 10^{-21}$cm$^3$/s that is orders of magnitude smaller than the hole capture coefficient $C_p = 1 \times 10^{-9}$cm$^3$/s, which has been observed both experimentally [16] and computationally [12].

Given such ambiguity in the trapping dynamics of the $C_N$ defect, a theoretical and experimental framework that aims to clarify the rate-limiting carrier capture step of a DARC and the initial charge state of the defect would be able to illuminate how recombination occurs at this defect. The model outlined in this work is a candidate for these studies, as it predicts that there will be an observable difference in ideality factor, depending on which carrier capture step is rate-limiting. Knowing which capture step is rate-limiting also allows for the determination of the initial charge state of the defect within a device, which is not always known. While experimental studies complementing this model are underway, it will be shown here that such an analysis of individual carrier capture steps is not possible with the conventional SNS analysis.

To do this, two p-i-n diodes that closely mimic MBE-grown diodes in structure, doping, and background impurity concentration are constructed. The p-, i-, and n-regions are all taken to be 300 nm in width, and both the net donor concentration in the n-region and the net acceptor concentration in the p-region are equal to $3 \times 10^{18}$ cm$^{-3}$. The background donor-like impurity concentration in the i-region is $5 \times 10^{15}$ cm$^{-3}$. In one diode a 15 nm carbon-doped plane is placed 40 nm from the n-side, while in the other diode the same plane is placed 40 nm from the p-side. The minority hole lifetime, $\tau_p$, in the carbon-doped plane was approximated as 1 ns while the minority electron lifetime, $\tau_n$, was approximated as 1 s; these approximate values roughly correspond to a carbon concentration of $5 \times 10^{17}$ cm$^{-3}$, using the capture coefficients computed in Reference [12] and the relation $\tau_{\{n,p\}} = \left(C_{\{n,p\}} \times N_t\right)^{-1}$. Since $\tau_n \gg \tau_p$, an impurity slab consisting of $C_N$ is more representative of the first model we derive where $\tau_n p \gg \tau_p n$ and should have ideality factors that follow those shown by moving the impurity plane in **Figure 5**. The free carrier concentrations throughout the diode can be computed using a 1-dimensional Schrodinger-Poisson drift-diffusion charge control solver [11]. Parameters used in the simulation are shown in **Table I**. Note that a 1% ionization efficiency for Mg at room temperature is used in the simulations, such that the dopant concentration [Mg]=$3 \times 10^{20}$ cm$^{-3}$ yields a net acceptor concentration ($N_A$) of approximately $3 \times 10^{18}$ cm$^{-3}$.

Device structures and their corresponding modeled J-V curves are shown in **Figure 7**. The solid blue curve gives the J-V curve modeled using the SNS analysis, adapted to a p-i-n diode from the p-n junction used in the original work [1]. The recombination current density as a function of applied voltage is given as

$$J_{R,\ SNS} = 2\left(\frac{kT}{qE}\right)\frac{q}{\tau_n \tau_p} n_i \left(\frac{qV}{2kT}\right), \tag{15}$$

with $E$ being the magnitude of the electric field at the plane of maximum recombination, computed using a 1-dimensional Schrodinger-Poisson drift-diffusion charge control solver [11]. Equation (14) was then combined with expression (4a) to give the total current density used in the plotted SNS J-V curve. The solid red curve plots the J-V characteristics using the model developed in this work, taking the sum of expressions (4a) and (12b) to compute the total current density. The drift-diffusion or "ideal" current density shown in expression (4a) was added to the recombination current density to account for drift-diffusion becoming the dominant mechanism of carrier transport as applied voltage approaches $V_{bi}$.

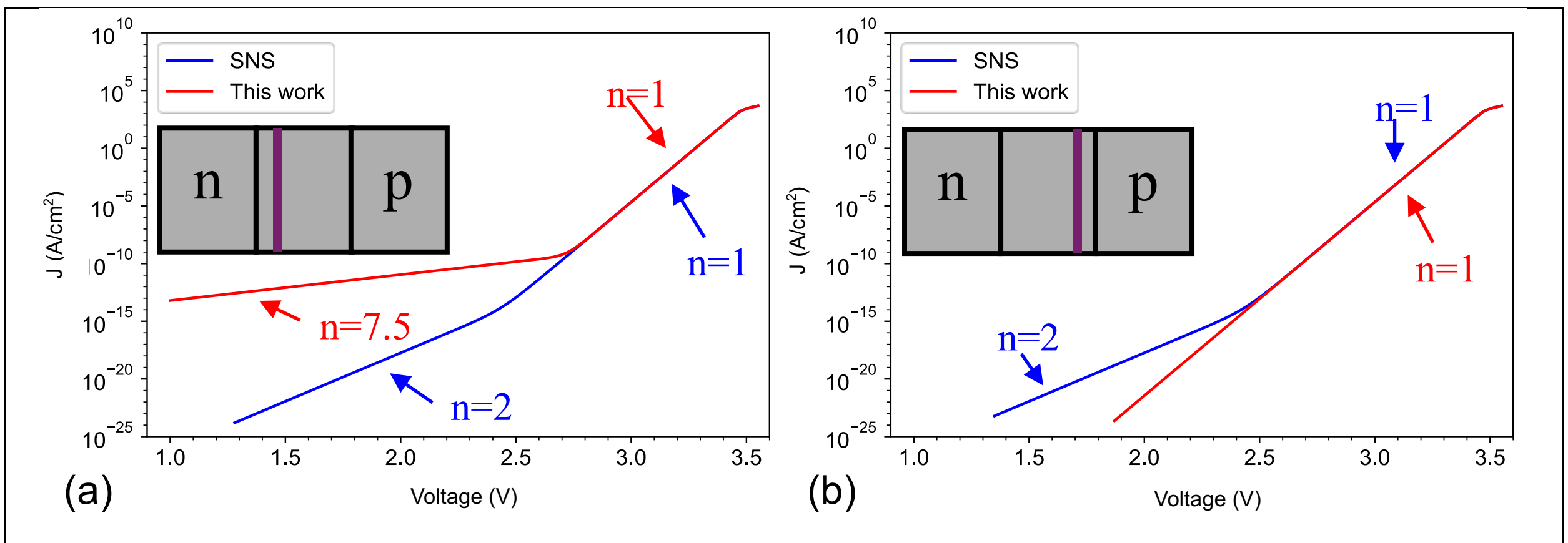

**Figure 7**: J-V curves computed for a p-i-n device with a C-doped plane near the (a) n-side or (b) p-side of the diode, using carrier lifetimes for SRH recombination $\tau_{SRH,n}$ and $\tau_{SRH,p}$. The blue curve is computed using the SNS model and the red curve is computed using the model developed in this work.

It is important to note that these computed J-V curves are not the same as those measured by experiment, as they neglect the effects of the current detection limit and series resistance. The current detection limit of any electrical measurement is the minimum current that can be measured and is usually in the tens or hundreds of microamperes. However, this effect does not have substantial impact on the analysis of **Figure 7** because they only affect data measured in the regions $V < 2.5$ V. Independent of these effects, substantial differences between the models appear when the carbon-doped plane is located 40 nm from the n-region, as is shown in panel (a). The J-V curve modeled with the SNS analysis is mainly dominated by the region with an ideality factor *n*=1, suggesting that current conduction is still dominated by drift-diffusion processes in experimentally accessible ranges of current density. However, the J-V predicted using our model shows substantial difference from the SNS analysis. Ideality factor below an applied bias of roughly 2.3 V was computed to be 7.5, far greater than the value of 2 expected using the standard SNS analysis if recombination is present in the depletion region. This is in line with observations of ideality factors in the range of 7-9 made by Fedison et al. [17], which they postulate to be due to the presence of deep-level states within the band gap. However, the current density throughout the *n*=7.5 region is small, leading to potential problems with experimental observation, though accounting for other modes of recombination can aid in this. Due to the experimentally imposed limits on J-V measurement, comparison of the models shown in panel (b) for the diode with a carbon-doped plane 40 nm from the p-region are essentially identical. Both models, in possessing an ideality factor of 1, predict that drift-diffusion processes will dominate at all experimentally accessible regions of the J-V curve, as they only begin to diverge at physically unmeaningful levels of current density. This calculation of the ideality factor exceeding the value of 2 when the impurity plane composed of $C_N$ acceptor traps is close to the n-side of the junction and ideality factor being 1 when the impurity plane composed of $C_N$ acceptor traps is close to the p-side of the junction is in excellent agreement with our model of electron capture-limited impurity planes shown in **Figure**

**5**. In both cases, the defects in the impurity plane are not behaving as recombination centers unlike what the conventional SNS model predicts. This is illustrative of how the new model captures recombination physics specific to wide-band gap semiconductors that are absent in the original SNS analysis.

However, since defect-assisted recombination does not only occur through thermally activated processes like MPE, the proposed mechanism by which SRH recombination happens, recombination via radiative capture processes should also be accounted for. In the case of the $C_N$ defect, the radiative electron capture coefficient $c_{rad,n}$ $(= 7 \times 10^{-14} \frac{\text{cm}^3}{\text{s}})$ is substantially larger than $c_{MPE,n}(= 1.58 \times 10^{-23} \frac{\text{cm}^3}{\text{s}})$, the electron capture coefficient for the MPE process. This implies that while electron capture via thermally assisted processes is improbable, radiative capture of electrons is much more likely. To account for both radiative and thermally assisted capture processes, an effective carrier lifetime can be computed using the expression

$$\frac{1}{\tau_{eff}} = \frac{1}{\tau_{rad}} + \frac{1}{\tau_{MPE}} = N_t C_{rad} + N_t C_{MPE}$$

Substituting $\tau_{eff,n}$ into (7b), new J-V curves can be generated that account for both radiative and thermally assisted carrier capture, shown in **Figure 8**. As ideality factor is only determined by the argument in the exponential term, the ideality factors shown in **Figure 8** are the same as those in **Figure 7** for their respective models. By contrast, carrier lifetimes impact overall current density as they only appear in the constant prefactor of the expressions, for both the SNS and new models. As lifetimes appear in the denominator of this prefactor for either model, shorter carrier lifetime leads to a larger recombination current density, as observed in **Figure 8** compared to **Figure 7**. As such, accounting for radiative carrier capture indicates that differences in ideality factor relates to the rate-limiting carrier capture mechanism for each capture step and can indicate which capture step is rate-limiting overall. It is interesting to note that electron and hole capture are each dominated by different capture mechanisms, rather than being somewhat equally affected by both MPE and radiative capture. For electrons, $\tau_{eff,n} = 2.9 \times 10^{-5}$ s $\approx \tau_{n,rad}$, with $\tau_{MPE,n}$ being essentially negligible. The opposite is true for hole capture, as $\tau_{eff,p} = 5.7 \times 10^{-9}$ s $\approx \tau_{\text{MPE,p}}$, indicating that radiative hole capture is unlikely compared to thermally assisted hole capture. This analysis indicates that for low current injection (where three-body recombination processes can be ignored), full DARCs can be completed through different carrier capture mechanisms, depending on which dominates for a particular carrier capture step.

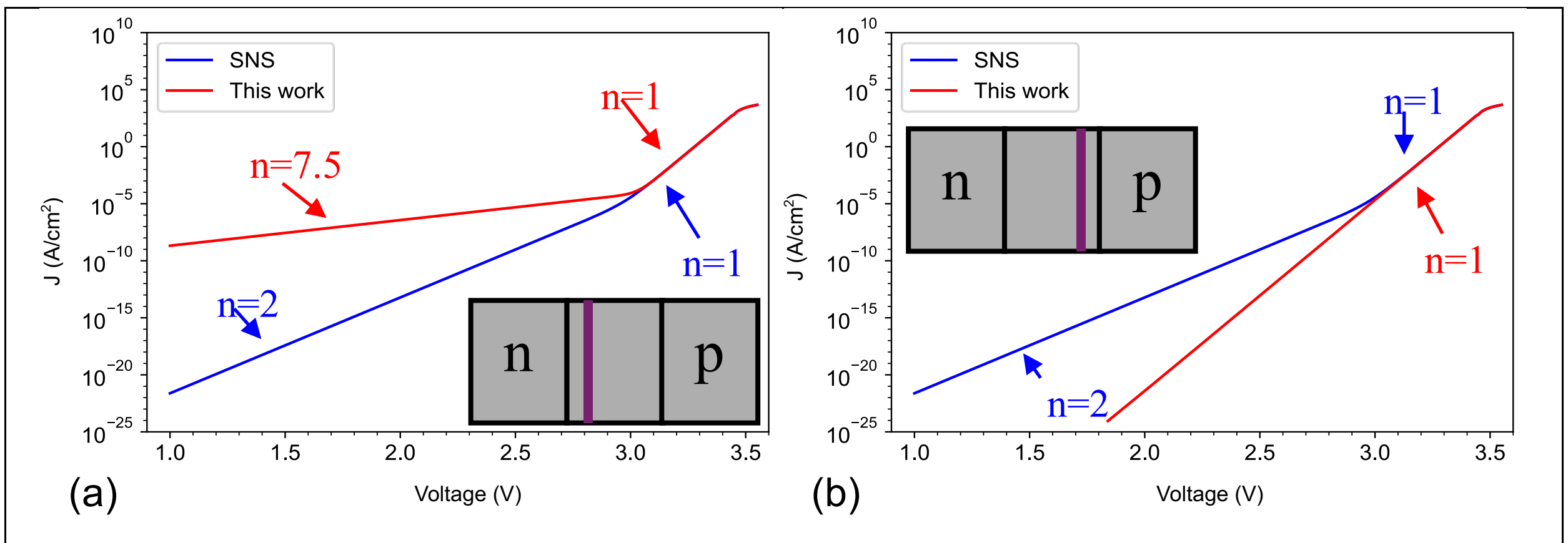

**Figure 8**: J-V curves computed for a p-i-n device with a C-doped plane near the (a) n-side or (b) p-side of the diode, using the effective carrier lifetimes $\tau_{eff,n}$ and $\tau_{eff,p}$. The blue curve is computed using the SNS model and the red curve is computed using the model developed in this work.

The observed differences in ideality factor depending on where in the depletion region the carbon-doped plane is located can illuminate how nonradiative recombination is taking place at the $C_N$ defect. For the case of the plane located closer to the n-region (**Figure 8a**), there is a substantial excess of free electrons compared to free holes in the location of the plane. Because the J-V curve and resultant ideality factor suggests that enhanced recombination is taking place within the depletion region of the device, implying that excess electrons are required to complete a DARC. This is exemplified by the reduced recombination in the depletion region shown by the J-V and ideality factor for the case of the carbon-doped plane being located close to the p-type region. In this case, a large excess of free holes compared to free electrons now exists in the region surrounding the impurity plane, implying that such a quantity of holes is not required to complete the recombination cycle. This indicates that electrons are the rate-limiting carrier capture step within the DARC. Thus, the model developed in this work illuminates the specific order of carrier capture at the $C_N$ defect.

### Case Study: Predicted J-V curves of p-i-n with i-region uniformly doped with C

Experimental observations of high-quality vertical GaN p-i-n structures grown by $NH_3$MBE [17] and MOCVD [18] have shown that the highest observed ideality factors are generally ~2 within a range of voltages of approximately 2-2.5V. This is in agreement with the original model SNS model but contrary to observations from studies involving p-n junctions [5], [17], LEDs [6], [7], and the impurity plane p-i-n diodes developed in the present work. Despite agreement with the SNS model, fundamental material differences between semiconductors of the 1950s (Si, Ge) and present-day wide-band gap semiconductors (GaN, SiC, $\beta$-$Ga_2O_3$) compels further study of diode characteristics for diodes fabricated from such wide-band gap materials. Namely, even with “clean” methods of epitaxial growth (e.g. MBE), low background levels of impurity elements like carbon, oxygen, and various transition metals can incorporate into epitaxially grown films with concentrations at least in the mid-$10^{15}$ $\text{cm}^{-3}$. Thus, in the present section, we consider how J-V characteristics for a GaN-based p-i-n diode are impacted by the presence of uniform carbon concentration of $5 \times 10^{15}$ $\text{cm}^{-3}$ throughout the depletion region, shown schematically in **Figure 9**. The diode considered here is identical to that considered in the prior section, with the only difference being the width of the carbon-doped region.

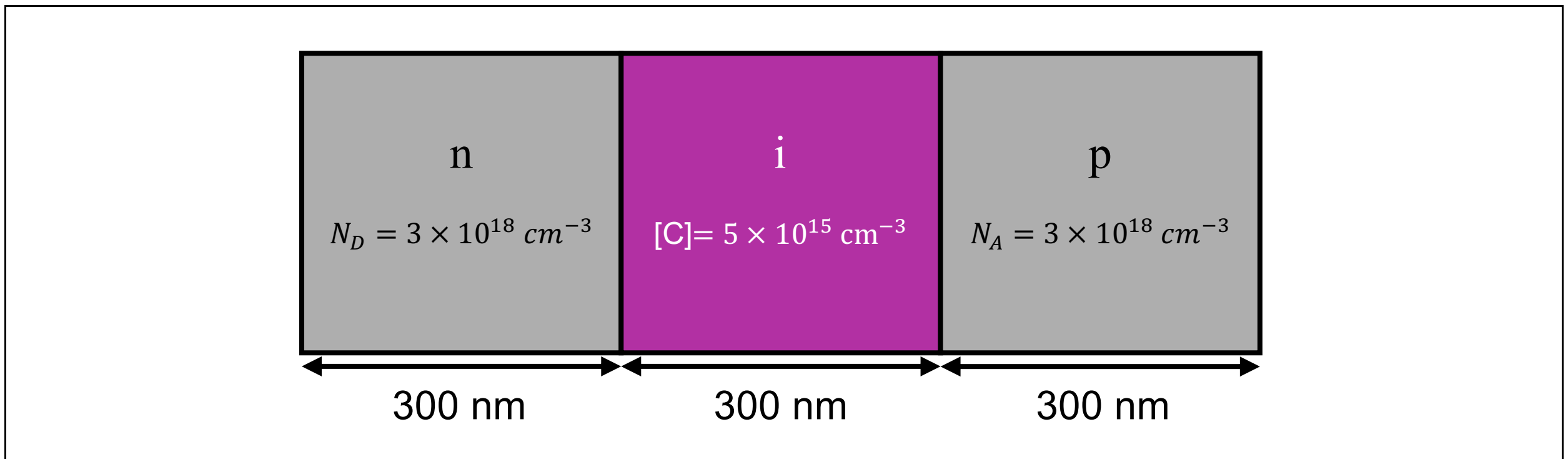

**Figure 9**: Schematic of p-i-n structure with uniform C-doping throughout the depletion region used in the following calculations.

Recall, the net recombination rate $R$ in its simplified form as it applies to GaN:

$$R \approx \frac{np - n_i^2}{\tau_n p + \tau_p n}$$

As done previously, the recombination-generation current $J_{RG}$ is obtained by integrating $R$ over the region of recombination and multiplying by the fundamental unit of charge

$$J_{RG} = q \int_{\frac{-W}{2}}^{\frac{W}{2}} \frac{np - n_i^2}{\tau_n p + \tau_p n} dx\ ,$$

where the origin is still defined at the center of the depletion region (which is approximately the same as the i-region width), so the integral is bounded between $\pm \frac{W}{2}$. Since the assumption that $n$ and $p$ are constant with $x$ can no longer be taken, the function must be expanded in terms of $x$ and then integrated. Substituting the expanded forms of $n$ and $p$, the equation becomes

$$J_{RG} = q \int_{\frac{-W}{2}}^{\frac{W}{2}} \frac{n_i \left[\exp\left(\frac{qV}{kT}\right) - 1\right] dx}{\tau_n \exp\left[\frac{q}{WkT}(V_{bi} - V)x - \frac{q}{2kT}(V_{bi} - V) + \frac{E_G}{2kT}\right] + \tau_p \exp\left[\frac{q}{WkT}(V - V_{bi})x + \frac{q}{2kT}(V + V_{bi}) - \frac{E_g}{2kT}\right]}$$

To condense the expression, the product of constants in the numerator can be pulled outside the integral while three new variables can be defined:

$$A \equiv \tau_n \exp\left[\frac{E_g - q(V_{bi} - V)}{2kT}\right], B \equiv \tau_p \exp\left[\frac{q(V + V_{bi}) - E_g}{2}\right], C \equiv \frac{q}{WkT}(V_{bi} - V)$$

Rewriting the expression and completing the integral, we get

$$J_{RG} = qn_i\left[\exp\left(\frac{qV}{kT}\right) - 1\right]\int_{\frac{-W}{2}}^{\frac{W}{2}} \frac{dx}{Ae^{Cx} + Be^{-Cx}} = qn_i\left[\exp\left(\frac{qV}{kT}\right) - 1\right]\cdot\frac{1}{C\sqrt{AB}}\tan^{-1}\left(\sqrt{\frac{A}{B}}e^{Cx}\right)\Bigg|_{-\frac{W}{2}}^{\frac{W}{2}}$$

The final answer is then obtained by substituting in the original values for $A, B,$ and $C$, and the recombination-generation current becomes

$$J_{RG} = \frac{WkTn_i\left[\exp\left(\frac{qV}{kT}\right) - 1\right]}{\sqrt{\tau_n\tau_p}\exp\left(\frac{qV}{2kT}\right)(V_{bi} - V)}\left\{\tan^{-1}\left[\sqrt{\frac{\tau_n}{\tau_p}}\exp\left(\frac{E_g - qV_{bi}}{2kT}\right)\exp\left(\frac{q(V_{bi} - V)}{2kT}\right)\right] - \tan^{-1}\left[\sqrt{\frac{\tau_n}{\tau_p}}\exp\left(\frac{E_g - qV_{bi}}{2kT}\right)\exp\left(\frac{-q(V_{bi} - V)}{2kT}\right)\right]\right\} \quad (16)$$

It is interesting to note that contrary to the earlier case of having the impurity-doped slab, the ideality factor for the uniform doping case no longer has a closed-form expression. Additionally, most of the physics is contained in the prefactor, as the bracketed difference will simply yield some quantity between 0 and $\frac{\pi}{2}$, inclusive.

Comparison of this model with the SNS analysis, as applied to the case of uniform carbon doping throughout the depletion region, is shown in **Figure 10**, using either $\tau_{MPE}$ or $\tau_{eff}$. While the SNS analysis tends to underpredict the recombination-generation current by roughly three orders of magnitude, the overall shape of the J-V characteristics given by either model is almost identical. For both cases, an ideality factor of $n = 2$ characterizes the recombination-generation current at lower voltages, while drift-diffusion processes dominate past ~2.2V ($\tau = \tau_{MPE}$) and ~2.5V ($\tau = \tau_{eff}$). This agrees with the observations made in Farzana *et al.* [17] and Ghosh *et al.* [18], implying that the ideality factor of $n = 2$ observed at lower biases for those GaN-based p-i-n diodes is likely due to background impurity incorporation during growth of the devices.

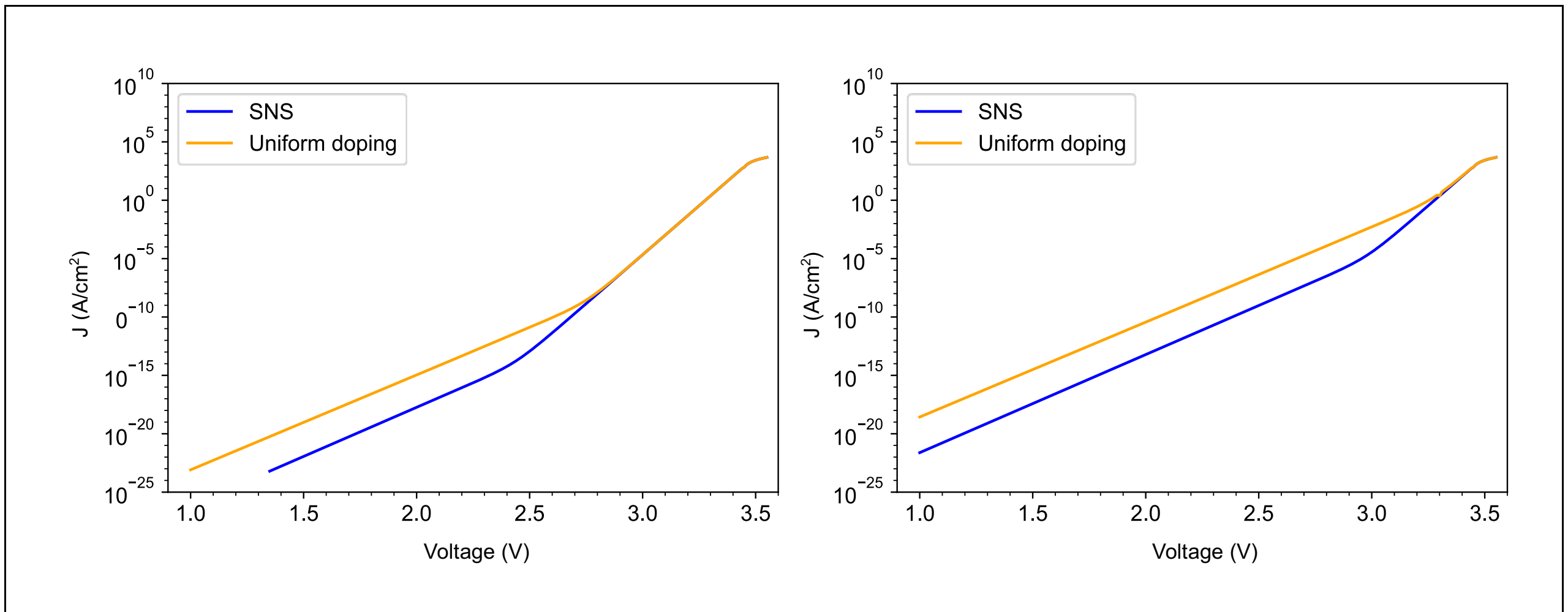

**Figure 10**: J-V characteristics for a diode with a uniformly carbon-doped i-region, computed using (a) $\tau = \tau_{MPE}$ and (b) $\tau = \tau_{eff}$. The blue curve is computed using the SNS model and the orange curve is computed using the model developed in this work.

This difference in ideality factor for the uniformly doped diode compared to the doped slab diode can be explained using arguments similar to those proposed by Sah *et al.* [1]. The factor of 2 in the denominator in the SNS analysis comes from the fact that the authors take $n = p$ as the point at which recombination is maximized, which vary as $n = p = n_i \exp\left(\frac{qV}{2kT}\right)$, since $np = n_i^2 \exp\left(\frac{qV}{kT}\right)$. In this work, the factor of 2 comes from boundary conditions used to define $E_{Fn} - E_i$ and $E_i - E_{Fp}$ in terms of known quantities. More specifically, since $E_i$ is used as a reference point, and $E_i \approx \frac{E_g}{2}$, 2 then appears in the denominator of the argument of the exponential. Though a closed-form expression for ideality factor could not be obtained for this case, the observed ideality factor of 2 becomes clear upon further analysis of the expression for $J_{RG}$. As already mentioned, the prefactor contains the core physics dictating the device's J-V behavior. Neglecting the 1 in the numerator of equation (16), the exponential terms in the numerator and denominator of the prefactor can be simplified into the factor $\exp\left(\frac{qV}{2kT}\right)$. There is also the factor of $\frac{1}{V_{bi}-V}$ in the same expression, but the linear variation here is negligible compared to the exponential variation $\exp\left(\frac{qV}{2kT}\right)$, so it can be neglected. Thus, the resulting ideality factor of 2 from the model devised in this work arises because the simplified factor of $\exp\left(\frac{qV}{2kT}\right)$ dominates over other voltage-dependent terms in the expression. The agreement in ideality factor between this work and the SNS model is therefore due to coincidence rather than any fundamental physical arguments shared between the two models.

## Discussion

This model illustrates that the diode ideality factor is only bounded by a lower limit of 1. The specific value of the ideality factor defined at a particular location within the diode will depend on where an impurity plane is placed within the diode's depletion region and the capture lifetimes of the defect that the impurity plane in composed of. This aspect of the model and device structure can be used to determine which carrier capture step is the slowest, or rate-limiting, step within the full recombination cycle. If hole capture is rate-limiting, then the ideality factor will diverge to a large value for an impurity slab placed near the i-p interface and will be 1 for an impurity slab placed at the i-n interface. By contrast, if electron capture is rate-limiting, then the ideality factor will diverge to a large value for an impurity slab placed near the i-n interface and will be 1 for an impurity slab placed near the i-p interface. These results are in agreement with those shown in Reference [7], despite that work using a full LED structure and neglecting the presence of traps in the active region. This indicates that this simplified structure adequately models an LED while removing other complexities not necessary for understanding core recombination physics.

In this way, this model and device structure can be employed to experimentally determine the rate-limiting carrier capture step of a recombination cycle in a straightforward manner, or the most favorable initial charge state of the defect. This would be the first ever electrical characterization of the individual carrier capture steps involved in a recombination cycle. By changing which impurity is used in the impurity plane, it will be possible to characterize the nonradiative recombination behavior of common impurities in wide band gap semiconductors such as GaN. This will yield a greater understanding of how certain common contaminants adversely affect wide band gap-based devices, as well as open the door for mitigating these deleterious effects through epi growth and device processing strategies.

Beyond utilizing the diode ideality factor to characterize individual recombination steps, the current density yielded by this model can give information regarding the carrier capture coefficients. As shown in expressions (7b) and (12b), the current density is inversely proportional to the minority carrier lifetimes $\tau_n$ or $\tau_p$, depending on which carrier capture step is rate-limiting. As discussed in [19], the minority carrier lifetimes are inversely proportional to the carrier capture coefficients, $c_n$ and $c_p$ for electrons and holes, respectively. While deep-level optical spectroscopy (DLOS) allows one to obtain carrier capture cross-sections, other quantities (like the average thermal velocity of carriers) are needed to arrive at the capture coefficients [20] described in detailed models of recombination, like in References [10] and [12]. While it is not impossible to compute, this is a more indirect and difficult way of obtaining carrier capture coefficients that is not necessarily accurate. However, using the model proposed in this work, it is much simpler and more straightforward to arrive at the carrier capture coefficients, as all other quantities in expressions (7b) and (12b) are known, besides $c_n$ or $c_p$.

Additionally, it is interesting to note that final expressions for total diode current density and ideality factor have no explicit dependence on impurity energy level. The factors $n_1$ and $p_1$ are the only quantities that depend on the defect energy level, but they were found to be negligible compared to the carrier concentrations *n* and *p*. This is true for even shallow defects (~200 meV from the band edges), as $n_1$ or $p_1$ would still be several orders of magnitude less than the carrier concentrations $n$ and $p$ in this case. The only parameter that is explicitly related to the specific impurity being considered is the minority carrier lifetimes, $\tau_n$ and $\tau_p$, which may be indirectly related to the defect level though it is presently undetermined. This is surprising, since defect energy levels have been known to affect net recombination rates, as pointed out by Shockley and Read [2]. Taking expression (1) with the appropriate substitutions for the net recombination rate, it can be easily seen that *R* is, in fact, maximized when the defect energy level coincides with the mid-gap energy level (this is essentially the same as the intrinsic energy level in GaN). Thus, it would be expected that if net recombination rate shows a dependence on defect energy level, then so should ideality factor. However, since the factors $n_1$ and $p_1$ are directly dependent on the intrinsic carrier concentration $n_i$ – which is in the magnitude of $10^{-10}$ $\text{cm}^{-3}$ in GaN, reflecting its wide bandgap – $n_1$ and $p_1$ will always be much smaller in GaN than they would for a narrow bandgap semiconductor like Si. As narrow bandgap materials like Si and Ge were commonly studied during the time of Shockley and Read's breakthrough paper, it is not surprising that conclusions drawn from those early studies may not have as much bearing on studies done considering modern wide bandgap materials like GaN.

**Conclusion**

A new model for computing ideality factor was derived for a GaN p-i-n junction with a thin plane of material intentional doped with known impurities. Closed-form analytical expressions for the ideality factor of the diode were derived based on a modified SRH and SNS analysis. In close agreement with other theoretical work on ideality factors of LEDs, our model predicts that ideality varies exponentially with distance across the depletion region. By changing where the impurity plane is located within the diode depletion region – either closer to the n-region or closer to the p-region – one can determine which carrier capture step in a defect-assisted recombination cycle (DARC) is rate-limiting by utilizing this model for ideality factor. This would enable observation and characterization of trap-assisted recombination in GaN not previously done.

**Acknowledgements**

Support for this work was provided by the Vannevar Buch Faculty Fellowship managed by the Office of Naval Research, Award No. N00014-25-1-2040; the Taiwan National Science and Technology Council (NSTC), Grants No. 112-2221-E-002-214-MY3 and No. 113-2124-M-002-013-MY3; the Office of Naval Research, Award No. N00014-22-1-2808 I.C.L. gratefully acknowledges additional support provided by U.S. Department of Education Graduate Assistance in Areas of National Need (GAANN) fellowship. Additional support for T.T. was provided by the Department of Defense's National Defense Science and Engineering Graduate (NDSEG) fellowship.

**Conflict of Interest Statement**

The authors have no conflict of interest to declare.

**Author Contributions**

**Iris Celupica-Liu:** Conceptualization (equal); Formal analysis (lead); Investigation (lead); Methodology (equal); Writing – original draft (lead); Writing – review and editing (equal). **Tanay Tak:** Conceptualization (equal); Formal analysis (equal); Methodology (equal); Writing – review and editing (equal). **Yuh-Renn Wu:** Conceptualization (equal); Formal analysis (equal); Software (lead); Writing – review and editing (equal). **James S. Speck:** Conceptualization (equal); Formal analysis (equal); Methodology (equal); Writing – review and editing (equal); Funding acquisition (lead); Supervision (lead).

**Data Availability Statement**

Data used in this article is available from the corresponding author upon reasonable request.